\documentclass[prl,preprint,showpacs,amsmath,amssymb]{revtex4}
\usepackage{graphicx}
\usepackage{enumerate}

\def \mate<#1|#2|#3>{\mbox{$\langle {#1}|\,{#2}\,|{#3}\rangle$}}
\newcommand{\nn}{\nonumber}
\def\dfrac#1#2{\displaystyle\frac{#1}{#2}}

\newcommand{\pslash}{p\kern-1ex /}
\newcommand{\qslash}{q\kern-1ex /}
\newcommand{\lslash}{l\kern-1ex /}
\newcommand{\sslash}{s\kern-1ex /}
\newcommand{\kaslash}{k_a\kern-2ex /}
\newcommand{\kbslash}{k_b\kern-2ex /}
\newcommand{\Dslash}{{\cal D}\kern-1.5ex /}

\newcommand{\tr}{{\rm tr}}

\newcommand{\beqa}{\begin{eqnarray}}
\newcommand{\eeqa}{\end{eqnarray}}

\begin{document}
\preprint{YITP-14-12}

\title{Pion masses in 2-flavor QCD with $\eta$ condensation}

\author{
Sinya Aoki$^{1,2}$ and
Michael Creutz$^{3}$
}

\affiliation{
$^1${Yukawa Institute for Theoretical Physics, Kyoto University, Kyoto 606-8502, Japan }\\
$^3${Center for Computational Sciences, University of Tsukuba, Ibaraki 305-8571, Japan}\\
$^3${Physics Department 510A, Brookhaven National Laboratory, Upton, NY 11973,USA\footnote{This manuscript has been authored under contract
    number DE-AC02-98CH10886 with the U.S.~Department of Energy.
    Accordingly, the U.S. Government retains a non-exclusive,
    royalty-free license to publish or reproduce the published form of
    this contribution, or allow others to do so, for U.S.~Government
    purposes. }
}
}

\begin{abstract}
We investigate some aspects of 2-flavor QCD with $m_u\not= m_d$ at
low-energy, using the leading order chiral perturbation theory
including anomaly effects.  While nothing special happens at $m_u=0$
for the fixed $m_d\not=0$, the neutral pion mass becomes zero at two
critical values of $m_u$, between which the neutral pion field
condenses, leading to a spontaneously CP broken phase, the so-called
Dashen phase. We also show that the "topological susceptibility" in
the ChPT diverges at these two critical points. We briefly discuss a
possibility that $m_u=0$ can be defined by the vanishing the
"topological susceptibility.  We finally analyze the case of
$m_u=m_d=m$ with $\theta=\pi$, which is equivalent to $m_u= - m_d=-m$
with $\theta = 0$ by the chiral rotation.  In this case, the $\eta$
condensation occurs at small $m$, violating the CP symmetry
spontaneously.  Deep in the $\eta$ condensation phase, three pions
become Nambu-Goldstone bosons, but they show unorthodox behavior at
small $m$ that $m_\pi^2 = O(m^2)$, which, however, is shown to be
consistent with the chiral Ward-Takahashi identities.
\end{abstract}
\maketitle

\section{Introduction}
One of possible solutions to the strong CP problem  is "massless up quark",  
where the $\theta$ term in QCD can be rotated away by the chiral rotation of up quark without affecting other part of the QCD action. This solution, unfortunately, seems to be ruled out by results from lattice QCD simulations\cite{Nelson:2003tb}.

In a series of papers\cite{Creutz:1995wf, Creutz:2003xu,
  Creutz:2003xc,Creutz:2005gb,Creutz:2013xfa}, however, one of the
present authors has argued that a concept of "massless up quark" is
ill-defined if other quarks such as a down quark are all massive,
since no symmetry can guarantee masslessness of up quark in this
situation due to the chiral anomaly. In addition, it has been also
argued that a neutral pion becomes massless at some negative value of
up quark mass for the positive down quark mass fixed, and beyond that
point, the neutral pion field condenses, forming a spontaneous CP
breaking phase, so-called a Dashen phase\cite{Dashen:1971aa}.
Furthermore, at the phase boundary, the topological susceptibility is
claimed to diverge due to the massless neutral pion, while it may
become zero at the would-be ``massless up quark" point.

The purpose of this letter is to investigate above properties of QCD
with non-degenerate quarks in more detail, using the chiral
perturbation theory (ChPT) with the effect of anomaly included as the
determinant term.  For simplicity, we consider the $N_f=2$ case with
$m_u\not=m_d$, but a generalization to an arbitrary number of $N_f$ is
straightforward with a small modification.  Our analysis explicitly
demonstrates the above-mentioned properties such as an absence of any
singularity at $m_u=0$ and the existence of the Dashen phase with the
appearance of a massless pion at the phase boundaries.  We further
apply our analysis to the case of $m_u=m_d=m$ with $\theta=\pi$, which
is equivalent to $m_u= - m_d=-m$ with $\theta = 0$ by the chiral
rotation. We show that, while $\eta$ condensation occurs, violating
the CP symmetry spontaneously, three pions become Nambu-Glodstone (NG)
bosons at $m=0$ deep in the $\eta$ condensation phase.  We also show a
unorthodox behavior at small $m$ that $m_\pi^2 = O(m^2)$, which is
indeed shown to be consistent with the chiral Ward-Takahashi
identities (WTI).

\section{Phase structure, masses and topological susceptibility}
The theory  we consider in this letter is given by 
\begin{eqnarray}
{\cal L} &=& \frac{f^2}{2}\tr\,  \left(\partial_\mu U \partial^\mu U^\dagger \right) -\frac{1}{2}\tr\, \left( M^\dagger U + U^\dagger M\right)\nn \\
& -& \frac{\Delta}{2}\left(\det U + \det U^\dagger\right), 
\end{eqnarray}
where $f$ is the pion decay constant, $M$ is a quark mass matrix, and
$\Delta$ is a positive constant giving an additional mass to an eta
meson. Differences between an ordinary ChPT and the above theory we
consider are the presence of the determinant term\footnote{Based on
  the large $N$ behavior, it is standard to use $(\log\det U)^2$ term
  to incorporate anomaly effects into ChPT\cite{Witten:1980sp,Rosenzweig:1979ay,Kawarabayashi:1980dp,Arnowitt:1980ne}.  Since
  we can determine the phase structure only numerically in this case,
  we instead employ $\det U$ term in our analysis, which leads to he
  phase structure determined analytically.  We have checked that the two
  cases lead to a qualitatively similar phase structure except at large
  quark masses.}, which breaks $U(1)$ axial symmetry, thus
representing the anomaly effect, and field $U \in U(N_f)$ instead of
$U\in SU(N_f)$.  We here ignore $\det U$ terms with derivatives for simplicity, since
they do not change our conclusions.  For $N_f=2$, without a loss of
generality, the mass term is taken as \beqa M &=& e^{i\theta} \left(
\begin{array}{cc}
 m_u & 0 \\
0 &  m_d \\
\end{array}
\right)
\equiv
e^{i\theta} 2 B \left( 
\begin{array}{cc}
m_{0u} & 0 \\
0 &  m_{0d} \\
\end{array}
\right) , \eeqa where $B$ is related to the magnitude of the chiral
condensate, $m_{0u,0d}$ are bare quark masses, and $\theta$ represents
the $\theta$ parameter in QCD. We consider that any explicit $F\tilde
F$ term in the action has been rotated into the mass matrix.  In the
most of our analysis, we take $\theta=0$, but an extension of our
analysis to $\theta\not=0$ is straight-forward.

\begin{figure}[tbh]
\begin{center}
\scalebox{0.25}{\includegraphics{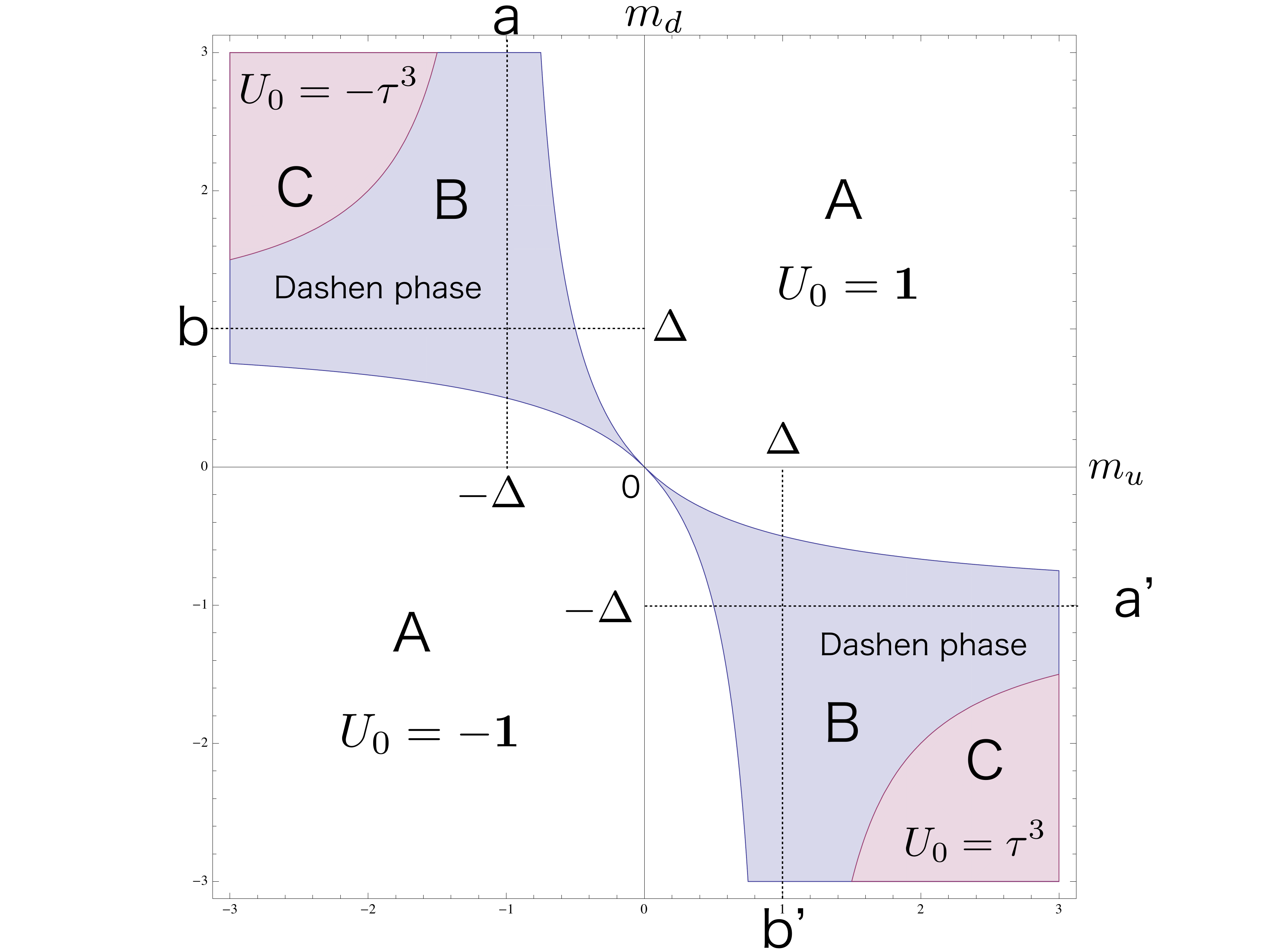}}
\end{center}
\caption{Phase structure in $m_u$-$m_d$ plain, where  the CP breaking Dashen phase are shaded in blue, while the CP preserving phase with $U_0=\tau^3$ (lower right) or $U_0=-\tau^3$ (upper left) are shaded in red. }
\label{fig:phase}
\end{figure}

Let us determine the vacuum structure of the theory at $m_u\not=
m_d$. Minimizing the action with \beqa U(x) &=& U_0
=e^{i\varphi_0}e^{i\sum_{a=1}^3 \tau^a\varphi_a} , \eeqa we obtain the
phase structure given in Fig.~\ref{fig:phase}, which is symmetric with
respect to $m_+\equiv m_u+m_d=0$ axis and $m_- \equiv m_d - m_u =0$
axis, separately.  The former symmetry is implied by the chiral
rotation that $U \rightarrow e^{i\pi \tau^1/2} U e^{i\pi \tau^1/2}$
($\psi \rightarrow e^{i\pi \gamma_5 \tau^1/2}\psi$ for the quark),
while the latter by the vector rotation that $U \rightarrow e^{i\pi
  \tau^1/2} U e^{-i\pi \tau^1/2}$ ($\psi \rightarrow e^{i\pi
  \tau^1/2}\psi$ )\footnote{This phase structure seems incompatible
  with the 1+1 flavor QCD with rooted staggered quarks, which is
  symmetric individually under $m_u\rightarrow - m_u$ or
  $m_d\rightarrow -m_d$. This suggests that the rooted trick for the
  staggered quarks can not be used in a region at $m_d m_u < 0$.}.

In the phase A (white), $U_0={\bf 1}_{2\times 2}$ (upper right) or
$U_0=-{\bf 1}_{2\times 2}$(lower left), while $U_0=\tau^3$ (lower
right) or $U_0=-\tau^3$ (upper left) in the phase C (shaded in red).
In the phase B (shaded in blue), we have nontrivial minimum with \beqa
\sin^2(\varphi_3) &=& \frac{ (m_d-m_u)^2\{(m_u+m_d)^2\Delta^2 -
  m_u^2m_d^2\}}{4m_u^3m_d^3} \\ \sin^2(\varphi_0) &=& \frac{
  (m_u+m_d)^2\Delta^2 - m_u^2m_d^2}{4m_u m_d\Delta^2}, \eeqa which
breaks CP symmetry spontaneously, since $\langle \pi^0\rangle= \tr\,
\tau^3 (U_0 - U_0^\dagger)/(2i) = 2\cos(\varphi_0)\sin(\varphi_3)$ and
$\langle \eta\rangle= \tr \, (U_0 - U_0^\dagger)/(2i)
=2\sin(\varphi_0)\cos(\varphi_3)$. This phase, where the neutral pion
and the eta fields condense, corresponds to the Dashen phase.  The
spontaneous CP breaking 2nd-order phase transition occurs at the
boundaries of the Dashen phase: Lines between phase A and phase B, on
which $\sin^2\varphi_3=\sin^2\varphi_0 = 0$, are defined by $
(m_d+m_u) \Delta + m_d m_u = 0$ (a line $\overline{a a^\prime}$) and
$(m_d+m_u) \Delta - m_d m_u = 0$ (a line $\overline{b b^\prime}$),
while those between $B$ and $C$, on which
$\sin^2\varphi_3=\sin^2\varphi_0 = 1$, are given by $(m_d-m_u)\Delta +
m_d m_u = 0$ (a line $\overline{ab}$) and $(m_d-m_u)\Delta - m_d m_u =
0$ (a line $\overline{a^\prime b^\prime}$). Note that
$\sin^2\varphi_3=1$ also on a $m_+=0$ line.

We next calculate pseudo-scalar meson masses in each phase.
Expanding $U(x)$ around $U_0$ as $ U(x) = U_0 e^{i \Pi(x)/f}$ with
\beqa
\Pi(x) &=& \left(
\begin{array}{cc}
\dfrac{\eta(x)+\pi_0(x)}{\sqrt{2}} & \pi_-(x) \\
\pi_+(x) & \dfrac{\eta(x)+\pi_0(x)}{\sqrt{2}} \\
\end{array}
\right),
\eeqa
the mass term is given by
\beqa
{\cal L}^{M} &=& \frac{m_+(\vec\varphi)}{4f^2}\left\{\eta^2(x) +\pi_0^2(x) + 2\pi_+(x)\pi_-(x)\right\}\nn \\
&+& \frac{\delta m}{2f^2}\eta^2(x) -\frac{m_-(\vec\varphi)}{2f^2}\eta(x)\pi_0(x),
\eeqa
where $m_\pm (\vec\varphi)=m_\pm \cos(\varphi_0)\cos(\varphi_3) + m_\mp \sin(\varphi_0)\sin(\varphi_3)$ 
with  $\delta m = 2 \Delta \cos(2\varphi_0)$. 
While the charged meson mass $m_{\pi_\pm}$ is simply given by
$m_{\pi_\pm}^2 = m_+(\vec\varphi)/(2f^2)$,  mass eigenstates,
\beqa
 \left(
\begin{array}{c}
\tilde\pi_0(x)  \\
\tilde\eta(x) \\
\end{array}
\right)  &=& 
\frac{1}{\sqrt{2X}}
 \left(
\begin{array}{c}
X_+^{1/2} \pi_0(x) + X_-^{1/2}\eta(x)  \\
X_-^{1/2}\pi_0(x) - X_+^{1/2}\eta(x) \\
\end{array}
\right) ,
\eeqa
have
\beqa
m_{\tilde \pi_0}^2 &=& \frac{1}{2f^2}\left[ m_+(\vec\varphi) +\delta m -X\right],\\
m_{\tilde\eta}^2 &=& \frac{1}{2f^2}\left[ m_+(\vec\varphi) +\delta m + X \right] ,
\eeqa
where $X=\sqrt{m_-(\vec\varphi)^2+\delta m^2}$ and $X_\pm =X\pm \delta m$.
We here choose $\tilde\pi_0$ and $\tilde\eta$ such that $m_{\tilde\pi_0}^2 \le m_{\tilde\eta}^2$.
It is then easy to see $m_{\tilde\pi_0}^2 \le m_{\pi_\pm}^2 \le m_{\tilde\eta}^2$.

By plugging $\varphi_0$ and $\varphi_3$ into the above formula, we obtain meson masses  in each phase.
Here we show  that $m_{\tilde\pi_0}^2=0$ at all phase boundaries, to demonstrate that the phase transition
is indeed of second order.   In the phase A, we have 
\beqa
m_{\tilde\pi_0}^2 &=& \frac{1}{2f^2}\left[ \vert m_+\vert +2\Delta -\sqrt{m_-^2+4\Delta}\right],
\eeqa
which becomes zero at  $(m_d+m_u)\Delta + m_d m_u =0$ (on $\overline{aa^\prime}$) and
at $(m_d+m_u)\Delta - m_d m_u =0$ (on $\overline{bb^\prime}$). 
Note that nothing special happens at $m_u=0$ (a massless up quark) at $m_d\not=0$ as
 $
m_{\tilde\pi_0}^2 =( \vert m_d\vert +2\Delta -\sqrt{m_d^2+4\Delta})/(2f^2)
$.
In the phase C, we obtain
\beqa
m_{\tilde\pi_0}^2 &=& \frac{1}{2f^2}\left[ \vert m_-\vert -2\Delta -\sqrt{m_+^2+4\Delta}\right],
\eeqa
$m_{\tilde\pi_0}^2=0$  at $(m_d-m_u)\Delta + m_d m_u = 0$ (on $\overline{ab}$) and at $(m_d-m_u)\Delta - m_d m_u = 0$ (on $\overline{a^\prime b^\prime}$). In addition, it is easy to check that the massless condition for $\tilde\pi_0$ that $m_+(\vec\varphi) + \delta m =\sqrt{m_-(\vec\varphi)^2+\delta m^2}$ in the phase B can be satisfied only on all boundaries of the phase B.

So far, we have shown three claims in \cite{Creutz:1995wf, Creutz:2003xu, Creutz:2003xc,Creutz:2013xfa} that (1) the Dashen phase with spontaneous CP breaking by the pion condensate exists
in non-degenerate 2-flavor QCD, (2) the massless neutral pion appears at the boundaries of the Dashen phase, and (3) nothing special happens at $m_u=0$ except at $m_d=0$.

We now consider the relation between the topological susceptibility
and $m_u$ in the ChPT.  To define the topological susceptibility in
ChPT, let us consider the chiral U(1) WTI given by \beqa
&&\langle\left\{ \partial^\mu A_\mu^0(x) + \tr \, M (U^\dagger(x) -
U(x) ) -2N_f q(x)\right\} {\cal O}(y) \rangle\nn
\\ &=&\delta^{(4)}(x-y)\langle \delta^0 {\cal O}(y)\rangle \eeqa where
$A_\mu =f^2\tr\{ U^\dagger(x)\partial_\mu U(x) - U(x)\partial
U^\dagger(x)\}$ is the U(1) axial current, ${\cal O}$ and $\delta^0
{\cal O}$ are an arbitrary operator and its infinitesimal local axial
U(1) rotation, respectively, and $2 N_f q(x) \equiv \Delta \{ \det
U(x) - \det U^\dagger(x) \}$ corresponds to the topological charge
density. Taking ${\cal O}(y)=q(y)$ and integrating over $x$, we define
the topological susceptibility in the ChPT through WTI as \beqa &&2N_f
\chi \equiv \int d^4 x \langle \{\partial^\mu A_\mu^0(x) + \tr\, M
(U^\dagger(x) - U(x) ) \} q(y) \rangle ,\nn \\ &&=
\frac{\Delta^2}{4}\int d^4 x \langle q(x) q(y) \rangle +
\frac{\Delta}{2} \langle \det U(x) + \det U^\dagger (x) \rangle ,
\eeqa where the second term comes from $\delta^0 q(x)$ in ChPT, which
is absent in QCD, but represent an effect of the contact term of $q(x)
q(y)$ in ChPT. The leading order in ChPT gives \beqa 2 N_f\chi &=& -
\frac{4\Delta^2
  m_+(\vec\varphi)}{m_+(\vec\varphi)^2-m_-(\vec\varphi)^2 +
  2m_+(\vec\varphi)\delta m }+\Delta .  \eeqa
 
 At $m_u=0$, we have $m_+(\vec\varphi) = m_-(\vec\varphi) = \vert m_d\vert $ and $\delta m = 2\Delta$, so that
 \beqa
 2N_f\chi = - 4\Delta^2 \vert m_d\vert /(4\vert m_d\vert\Delta) +\Delta = 0, 
 \eeqa
 which confirms the statement that (4) $\chi=0$ at $m_u=0$.   Since the denominator of $\chi$ is proportional to $m_{\tilde\pi_0}^2 \times m_{\tilde\eta}^2$,
 $\chi\rightarrow - \infty$ on all phase boundaries since $m_ {\tilde\pi_0}^2 =0$ and $m_+(\vec\varphi) > 0$,
 which again confirms the statement that (5) $\chi$ negatively diverges at the phase boundaries where
 the neutral pion becomes massless.
 
 We have confirmed the five statements in Ref.~\cite{Creutz:1995wf,
   Creutz:2003xu, Creutz:2003xc,Creutz:2013xfa}, (1) -- (5) in the
 above, by the ChPT analysis.  In addition, we have found a new CP
 preserving phase, the phase C, which has $U_0=\pm\tau^3$ instead of
 $U_0=\pm{\bf 1}_{2\times 2}$ of the phase A.  Since the phase C
 occurs at rather heavy quark masses such that $m_{u,d} = 2B
 m^0_{u,d}=O(\Delta)$, however, the leading order ChPT analysis may
 not be reliable for the phase C.  Indeed, the phase C seems to
 disappear if $(\log\det U)^2$ is employed instead of $\det U$.  Other
 properties, (1) -- (5), on the other hand, are robust, since they
 already occur near the origin ($m_u=m_d=0$) in the $m_u-m_d$ plain
 and they survive even if $(\log\det U)^2$ is used.
 
 The property (4) suggests an interesting possibility that one can
 define $m_u=0$ at $m_d\not=0$ in 2-flavor QCD from a condition that
 $\chi=0$. This is different from the standard statement that the
 effect of $\theta$ term is rotated away at $m_u=0$. We instead define
 $m_u=0$ from $\chi =0$, which is equivalent to an absence of the
 $\theta$ dependence if higher order cumulants of topological charge
 fluctuations are all absent.  A question we may have is whether
 $\chi=0$ is a well defined condition or not.  As already discussed in
 Ref.~\cite{Creutz:1995wf, Creutz:2003xu,
   Creutz:2003xc,Creutz:2013xfa}, a value of $\chi$, and thus the
 $\chi=0$ condition, depend on its definition at finite lattice
 spacing (cut-off). Although one naively expect such ambiguity of
 $\chi$ disappears in the continuum limit, we must check a uniqueness
 of $\chi$ explicitly in lattice QCD calculations by demonstrating
 that $\chi$ from two different definitions but at same physical
 parameters agree in the continuum limit. If the uniqueness of $\chi$
 can be established, one should calculate $\chi$ at the physical point
  of 1+ 1+1 flavor QCD in the continuum limit.  If $\chi\not=0$
 in the continuum limit, the solution to the U(1) problem by the
 massless up quark ( $\chi=0$ in our definition) is ruled out.

\section{Degenerate 2-flavor QCD at $\theta=\pi$}
In the remainder of this letter, as an application of our analysis, we
consider the 2-flavor QCD with $m_u=m_d=m$ and $\theta =0$, which is
equivalent to the 2-flavor QCD with $m_u = - m_d$ but $\theta = 0$. In
both systems, we have a SU(2) symmetry generated by $\{
\tau^1,\tau^2,\tau^3\}$ for the former or $\{
\tau^1\gamma_5,\tau^2\gamma_5,\tau^3 \}$ for the latter. We here give
results for the former case, but a reinterpretation of results in the
latter case is straightforward.

The vacuum is given by $\varphi_3 = 0$ and
\beqa
\cos\varphi_0 &=& \left\{
\begin{array}{lc}
1, &  2\Delta \le m \\
\dfrac{m}{2\Delta} ,  & -2\Delta < m < 2\Delta \\
-1,  & m \le -2\Delta \\
\end{array}
\right. ,
\eeqa
which leads to 
\beqa
\langle \bar\psi i\gamma_5 \psi \rangle  &=&
\left\{
\begin{array}{lc}
0, &  m^2 \ge 4\Delta^2 \\
\pm 2\sqrt{1-\dfrac{m^2}{4\Delta^2}},  &  m^2 < 4\Delta^2 \\
\end{array}
\right. , \\
\langle \bar\psi \psi \rangle &=&\left\{
\begin{array}{lc}
2, &  2\Delta \le m \\
\dfrac{m}{\Delta} ,  & -2\Delta < m < 2\Delta \\
-2,  & m \le -2\Delta \\
\end{array}
\right. , 
\eeqa
showing the spontaneous CP symmetry breaking at $m^2 < 4 \Delta^2$.
Note that $\langle \bar\psi \psi\rangle^2+ \langle \bar\psi i\gamma_5\psi\rangle^2=4$ at all $m$.

\begin{figure}[tbh]
\begin{center}
\scalebox{0.23}{\includegraphics{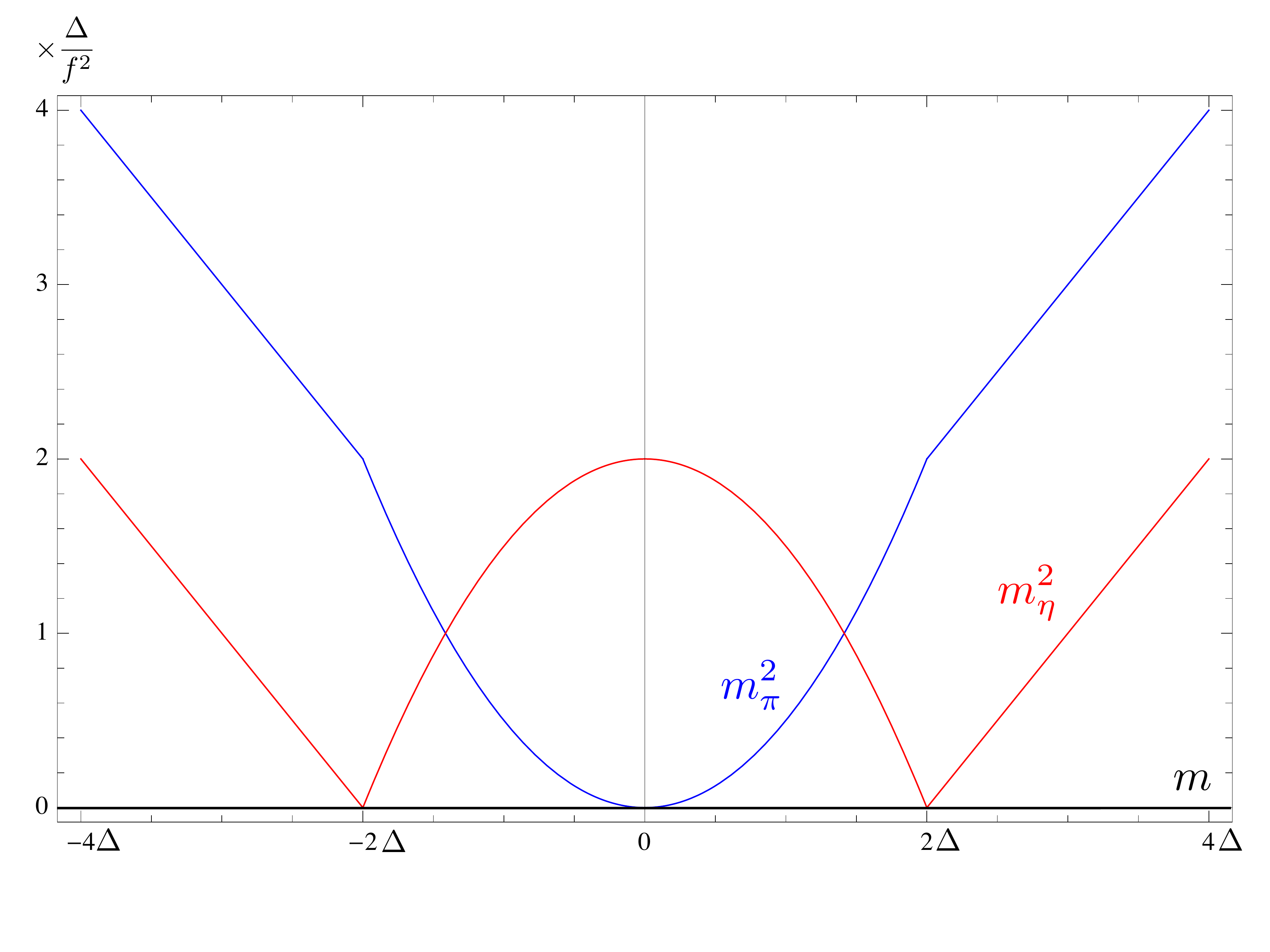}}
\end{center}
\caption{$m_\pi^2$ (blue) and $m_\eta^2$ (red) in unit of $\frac{\Delta}{f^2}$ as a function of $m$.}
\label{fig:PSmesons}
\end{figure}

PS meson masses are calculated as
\beqa
m_\pi^2&=& m_{\pi_\pm}^2 = m_{\pi_0}^2 = \left\{
\begin{array}{ll}
\dfrac{1}{2f^2} 2\vert m\vert, & m^2 \ge 4\Delta^2 \\
\dfrac{1}{2f^2}\frac{m^2}{\Delta}, & m^2 < 4\Delta^2 \\  
\end{array}
\right. ,\\
m_\eta^2 &=&  \left\{
\begin{array}{ll}
\dfrac{1}{2f^2} \left[2\vert m\vert-4\Delta \right], & m^2 \ge 4\Delta^2 \\
\dfrac{1}{2f^2}\frac{4\Delta^2 - m^2}{\Delta}, & m^2 < 4\Delta^2 \\  
\end{array}
\right. ,
\eeqa
where $\eta$ becomes massless at the phase boundaries at $m^2=4\Delta^2$, showing that $\eta$ is the massless mode associated with the spontaneous CP symmetry breaking phase transition,  while
three pion become massless Nambu-Goldstone modes at $m=0$. Fig.~\ref{fig:PSmesons} represents these behaviors.

As mentioned before, although ChPT analysis around the phase
transition points at $m^2 = 4 \Delta^2$ may not be
reliable\footnote{Indeed, both CP preserving phase, which correspond
  to the phase C, and the phase transition points, disappear if
  $(\log\det U)^2$ term is employed instead of $\det U$ term. In this
  case CP is broken at all $m$.}, we can trust the results near $m=0$
that the CP symmetry is spontaneously broken by the $\eta$
condensation in the degenerate 2-flavor QCD with $\theta=\pi$ and
three pions become massless NG bosons at $m=0$.  Pion masses, however,
behaves as $m_\pi^2 = m^2/(2f^2\Delta)$ near $m=0$, contrary to the
orthodox PCAC relation that $m_\pi^2 = \vert m\vert
/(2f^2)$\footnote{A similar behavior has been predicted in a different
  context\cite{Knecht:1994zb,Stern:1997ri,Stern:1998dy}.}. Let us show
that this unorthodox relation can be explained by the WTI. The
integrated WTI for the non-singlet chiral rotation with $\tau^3$ and
${\cal O}=\tr\, \tau^3 (U^\dagger - U)$ reads \beqa && m \int d^4x\,
\tr\,\tau^3 ( U^\dagger - U)(x) \tr\,\tau^3 ( U^\dagger - U)(y)\rangle
\nn \\ &=& -2\langle \tr\, (U+U^\dagger)(y)\rangle, \eeqa which leads
to \beqa m_{\pi_0}^2 &=& \frac{m}{f^2} \cos\varphi_0 =
\frac{m^2}{f^2}\frac{m}{2\Delta} .  \eeqa This tells us that one $m$
explicitly comes from the WTI, the other $m$ from the VEV of $\bar\psi
\psi$, giving the unorthodox relation that \beqa m_\pi^2 &=&
\frac{m^2}{2f^2\Delta} .  \eeqa It is interesting and challenging,
because of sign problems, to confirm this prediction by lattice QCD
simulations with $\theta=\pi$, and to consider possible applications
of this to particle physics \cite{Dashen:1971aa}. \\
 
\begin{acknowledgments}
S.A thanks Dr. T. Hatsuda  for useful comments.
S.A  is partially supported by Grant-in-Aid
for Scientific Research on Innovative Areas(No.2004:20105001) and
for Scientific Research (B) 25287046 and SPIRE (Strategic Program for Innovative REsearch).
\end{acknowledgments}

 \end{document}